\documentstyle[aps,preprint,floats,epsf,epsfig]{revtex}
\begin{document}
\def\be{\begin{eqnarray}}
\def\en{\end{eqnarray}}
\def\non{\nonumber}
\def\la{\langle}
\def\ra{\rangle}
\def\nc{N_c^{\rm eff}}
\def\vp{\varepsilon}
\def\a{{\cal A}}
\def\B{{\cal B}}
\def\c{{\cal C}}
\def\d{{\cal D}}
\def\e{{\cal E}}
\def\p{{\cal P}}
\def\t{{\cal T}}
\def\up{\uparrow}
\def\dw{\downarrow}
\def\vma{{_{V-A}}}
\def\vpa{{_{V+A}}}
\def\smp{{_{S-P}}}
\def\spp{{_{S+P}}}
\def\J{{J/\psi}}
\def\ov{\overline}
\def\Lqcd{{\Lambda_{\rm QCD}}}
\def\pr{{\sl Phys. Rev.}~}
\def\prl{{\sl Phys. Rev. Lett.}~}
\def\pl{{\sl Phys. Lett.}~}
\def\np{{\sl Nucl. Phys.}~}
\def\zp{{\sl Z. Phys.}~}
\def\lsim{ {\ \lower-1.2pt\vbox{\hbox{\rlap{$<$}\lower5pt\vbox{\hbox{$\sim$}
}}}\ } }
\def\gsim{ {\ \lower-1.2pt\vbox{\hbox{\rlap{$>$}\lower5pt\vbox{\hbox{$\sim$}
}}}\ } }

\font\el=cmbx10 scaled \magstep2{\obeylines\hfill December, 2002}

\vskip 1.5 cm

\centerline{\large\bf Comments on the Quark Content of the Scalar
Meson $f_0(1370)$}
\bigskip
\centerline{\bf Hai-Yang Cheng}
\medskip
\centerline{Institute of Physics, Academia Sinica}
\centerline{Taipei, Taiwan 115, Republic of China}
\medskip

\bigskip
\bigskip
\centerline{\bf Abstract}
\bigskip
{\small Based on the measurements of $(D_s^+,D^+)\to
f_0(1370)\pi^+$ we determine, in a model independent way, the
allowed $s\bar s$ content in the scalar meson $f_0(1370)$. We find
that, on the one hand, if this isoscalar resonance is a pure
$n\bar n$ state [\,$n\bar n\equiv(u\bar u+d\bar d)/\sqrt{2}\,]$, a
very large $W$-annihilation term will be needed to accommodate
$D_s^+\to f_0(1370)\pi^+$. On the other hand, the $s\bar s$
component of $f_0(1370)$ should be small enough to avoid excessive
$D_s^+\to f_0(1370)\pi^+$ induced from the external $W$-emission.
Measurement of $f_0(1370)$ production in the decay $D_s^+\to
K^+K^-\pi^+$ will be useful to test the above picture. For the
decay $D^0\to f_0(1370)\ov K^0$ which is kinematically barely or
even not allowed, depending on the mass of $f_0(1370)$, we find
that the finite width effect of $f_0(1370)$ plays a crucial role
on the resonant three-body decay $D^0\to f_0(1370)\ov
K^0\to\pi^+\pi^-\ov K^0$.

}

\pagebreak

\section{Introduction}
It is known that the identification of scalar mesons is difficult
experimentally and the underlying structure of scalar mesons is
not well established theoretically (for a review, see e.g.
\cite{Spanier,Godfrey,Close}). It has been suggested that the
light scalars below or near 1 GeV--the isoscalars $\sigma(500)$,
$f_0(980)$, the isodoublet $\kappa$ and the isovector
$a_0(980)$--form an SU(3) flavor nonet, while scalar mesons above
1 GeV, namely, $f_0(1370)$, $a_0(1450)$, $K^*_0(1430)$ and
$f_0(1500)/f_0(1710)$, form another nonet. A consistent picture
\cite{Close} provided by the data  suggests that the scalar meson
states above 1 GeV can be identified as a $q\bar q$ nonet with
some possible glue content, whereas the light scalar mesons below
or near 1 GeV form predominately a $qq\bar q\bar q$ nonet
\cite{Jaffe,Alford} with a possible mixing with $0^+$ $q\bar q$
and glueball states. This is understandable because in the $q\bar
q$ quark model, the $0^+$ meson has a unit of orbital angular
momentum and hence it should have a higher mass above 1 GeV. On
the contrary, four quarks $q^2\bar q^2$ can form a $0^+$ meson
without introducing a unit of orbital angular momentum. Moreover,
color and spin dependent interactions favor a flavor nonet
configuration with attraction between the $qq$ and $\bar q\bar q$
pairs. Therefore, the $0^+$ $q^2\bar q^2$ nonet has a mass near or
below 1 GeV.

As the quark content of $a_0(1450)$ and $K^*_0(1430)$ is quite
obvious, the internal structure of the isoscalars $f_0(1370)$,
$f_0(1500)$ and $f_0(1710)$ in the same nonet is controversial and
less clear. Though it is generally believed that $f_0(1370)$ is
mainly $n\bar n\equiv (u\bar u+d\bar d)/\sqrt{2}$, the content of
$f_0(1500)$ and $f_0(1710)$ still remains confusing. For example,
it has been advocated that $f_0(1710)$ is mainly $s\bar s$ and
$f_0(1500)$ mostly gluonic (see e.g. \cite{Amsler}), while the
analysis in \cite{Kleefeld} suggests a dominantly $s\bar s$
interpretation of $f_0(1500)$. How much is the fraction of glue in
each isoscalar meson is another important but unsettled issue.

Three-body decays of heavy mesons provide a rich laboratory for
studying the intermediate state resonances. The Dalitz plot
analysis is a powerful technique for this purpose. Many scalar
meson production measurements in charm decays are now available
from the dedicated experiments conducted at CLEO, E791, FOCUS, and
BaBar. The study of three-body decays of charmed mesons not only
opens a new avenue to the understanding of the light scalar meson
spectroscopy, but also enables us to explore the quark content of
scalar resonances.  In \cite{ChengDSP} we have studied the
nonleptonic weak decays of charmed mesons into a scalar meson and
a pseudoscalar meson. The scalar resonances under consideration
there are $\sigma$ [or $f_0(600)$], $\kappa$, $f_0(980)$,
$a_0(980)$ and $K^*_0(1430)$.

In this work we would like to explore the quark content of
$f_0(1370)$ from hadronic charm decays. Since $\rho\rho$ and
$4\pi$ are its dominant decay modes \cite{PDG}, it is clear that
$f_0(1370)$ is mostly $n\bar n$. However, how much is the $s\bar
s$ component allowed in the wave function of this isoscalar
resonance remains unknown. It turns out that the decay $D_s^+\to
f_0(1370)\pi^+$ is very useful for this purpose. If $f_0(1370)$ is
purely a $n\bar n$ state, it can proceed only via the
$W$-annihilation diagram. In contrast, if $f_0(1370)$ has an
$s\bar s$ content, the decay $D_s^+\to f_0(1370)\pi^+$ will
receive an external $W$-emission contribution. Therefore, this
mode is ideal for determining the $s\bar s$ component in
$f_0(1370)$.

We would work in the model-independent quark-diagram approach in
which a least model-independent analysis of heavy meson decays can
be carried out. In this diagrammatic scenario, all two-body
nonleptonic weak decays of heavy mesons can be expressed in terms
of six distinct quark diagrams \cite{Chau,CC86,CC87}: $T$, the
color-allowed external $W$-emission tree diagram; $C$, the
color-suppressed internal $W$-emission diagram; $E$, the
$W$-exchange diagram; $A$, the $W$-annihilation diagram; $P$, the
horizontal $W$-loop diagram; and $V$, the vertical $W$-loop
diagram. (The one-gluon exchange approximation of the $P$ graph is
the so-called ``penguin diagram".) It should be stressed that
these quark diagrams are classified according to the topologies of
weak interactions with all strong interaction effects included and
hence they are {\it not} Feynman graphs. Therefore, topological
graphs can provide information on final-state interactions (FSIs).

Based on  SU(3) flavor symmetry, this model-independent analysis
enables us to extract the topological quark-graph amplitudes and
see the relative importance of different underlying decay
mechanisms. For $D\to SP$ decays ($S$: scalar meson, $P$:
pseudoscalar meson), there are several new features. First, one
can have two different external $W$-emission and internal
$W$-emission diagrams, depending on whether the emission particle
is a scalar meson or a pseudoscalar one. We thus denote the prime
amplitudes $T'$ and $C'$ for the case when the scalar meson is an
emitted particle \cite{ChengDSP}. Second, because of the smallness
of the decay constant of the scalar meson (see e.g. \cite{Diehl}),
it is expected that $|T'|\ll |T|$ and $|C'|\ll |C|$. Moreover, in
flavor SU(3) limit, the primed amplitudes $T'$ and $C'$ diminish
under the factorization approximation due to the vanishing decay
constants of scalar mesons \cite{ChengDSP}. Third, since the
scalar mesons $f_0(1370)$, $a_0(1450)$, $K^*_0(1430)$,
$f_0(1500)/f_0(1710)$ and the light ones
$\sigma,~\kappa,~f_0,~a_0$ fall into two different nonets, one
cannot apply SU(3) symmetry to relate the topological amplitudes
in $D^+\to f_0(1370)\pi^+$ to, for example, those in $D^+\to
f_0(980)\pi^+$.

The reduced quark-graph amplitudes $T,C,E,A$ for Cabibbo-allowed
$D\to PP$ decays have been extracted from the data with the
results \cite{Rosner}:
 \be \label{PP}
 T &=& (2.67\pm0.20)\times 10^{-6}\,{\rm GeV}, \non \\
 C &=& (2.03\pm0.15)\,{\rm Exp}[-i(151\pm4)^\circ]\times
 10^{-6}\,{\rm GeV}, \non \\
 E &=& (1.67\pm0.13)\,{\rm Exp}[\,i(115\pm5)^\circ]\times 10^{-6}\,{\rm GeV},
 \non \\ A &=& (1.05\pm0.52)\,{\rm Exp}[-i(65\pm30)^\circ]\times 10^{-6}\,{\rm
 GeV}.
 \en
These  amplitudes will be employed as a guidance when we come to
discuss $D\to f_0(1370)P$ decays below.

\section{quark content of \lowercase{$f_0(1370)$}}
The mass and width of the isoscalar resonance $f_0(1370)$ are far
from being well established. The recent study of $f_0(1370)$
production in $pp$ interactions by WA102 \cite{WA102} yields a
mass of order 1310 MeV and width of order $100-250$ MeV (see
\cite{WA102} for the detailed values of the mass and width). The
E791 experiment by analyzing $D_s^+\to\pi^+\pi^+\pi^-\to
f_0(1370)\pi^+$ gives a higher mass of $1434\pm18\pm9$ MeV and
width of $172\pm32\pm6$ MeV \cite{E791Ds}. The mass and width
quoted by the Particle Data Group \cite{PDG} span a wide range,
namely, $m_{f_0(1370)}=1200-1500$ MeV and
$\Gamma_{f_0(1370)}=200-500$ MeV.

Since $\rho\rho$ and $4\pi$ are the dominant decay modes of
$f_0(1370)$ \cite{PDG}, it is clear that this isoscalar resonance
is predominately $n\bar n$. In the present work we would like to
study its content from the three-body decays of charmed mesons to
see how much the $s\bar s$ component is allowed in $f_0(1370)$.

The production of the resonance $f_0(1370)$ in hadronic decays of
charmed mesons has been observed in the decay $D^0\to \ov
K^0\pi^+\pi^-\to f_0(1370)\ov K^0$ by ARGUS \cite{ARGUS}, E687
\cite{E687} and CLEO \cite{CLEO}, in $D_s^+\to\pi^+\pi^+\pi^-\to
f_0(1370)\pi^+$ by E791 \cite{E791Ds}, in $D^+\to K^+K^-\pi^+\to
f_0(1370)\pi^+$ by FOCUS \cite{FOCUS} and in
$D^+\to\pi^+\pi^-\pi^+\to f_0(1370)\pi^+$ by E791 \cite{E791Dp},
respectively, with the results
 \be \label{BRdata}
 \B(D^0\to f_0(1370)\ov K^0)\B(f_0(1370)\to\pi^+\pi^-) &=& \cases{
 (4.7\pm1.4)\times 10^{-3} & ARGUS,E687 \cr (5.9^{+1.8}_{-2.7})
 \times 10^{-3} & CLEO} \non \\
 \B(D^+\to f_0(1370)\pi^+)\B(f_0(1370)\to K^+K^-) &=&
 (6.2\pm 1.1)\times 10^{-4}\qquad{\rm FOCUS}   \\
 \B(D^+\to f_0(1370)\pi^+)\B(f_0(1370)\to \pi^+\pi^-) &=&
 (7.1\pm 6.4)\times 10^{-5}\qquad{\rm E791}  \non \\
 \B(D^+_s\to f_0(1370)\pi^+)\B(f_0(1370)\to\pi^+\pi^-) &=&
 (3.3\pm 1.2)\times 10^{-3}\qquad{\rm E791} \non
 \en
However, the E791 measurement of $D^+\to f_0(1370)\pi^+$ does not
have enough statistic significance and hence we will ignore it in
the ensuing discussion. The branching fractions of $f_0(1370)$
into $\pi^+\pi^-$ and $K^+K^-$ are unknown, though several early
attempts have been made (see \cite{PDG}).

We write the general $f_0(1370)$ flavor wave function as
 \be \label{f0w.f.}
 f_0(1370)=n\bar n \cos\theta+s\bar s\sin\theta.
 \en
In terms of the quark-diagram amplitudes depicted in Fig. 1, the
decay amplitudes of $D\to f_0(1370)P$ have the expressions
 \be \label{qda}
 A(D^+\to f_0(1370)\pi^+) &=&
 V_{cd}V_{ud}^*(T_d+A_{u,d})+V_{cs}V_{us}^*C'_s,  \non \\
 A(D^0\to f_0(1370)\ov K^0) &=&
 V_{cs}V_{ud}^*(C_u+E_{d,s}),  \\
 A(D^+_s\to f_0(1370)\pi^+) &=&
 V_{cs}V_{ud}^*(T_s+A_{u,d}), \non
 \en
where the subscript $q$ of the topological amplitude denotes the
$q\bar q$ component of $f_0(1370)$ involved in its production. In
terms of the mixing angle $\theta$ defined in Eq. (\ref{f0w.f.})
we have $T_s=\sqrt{2}\,T_d\tan\theta$. We see that if $f_0(1370)$
is a $n\bar n$ state in nature, the decay $D^+_s\to
f_0(1370)\pi^+$ can only proceed through the topological
$W$-annihilation diagram.

\begin{figure}[t]
\hspace{2.5cm}
%\hskip 2.5cm
  \psfig{figure=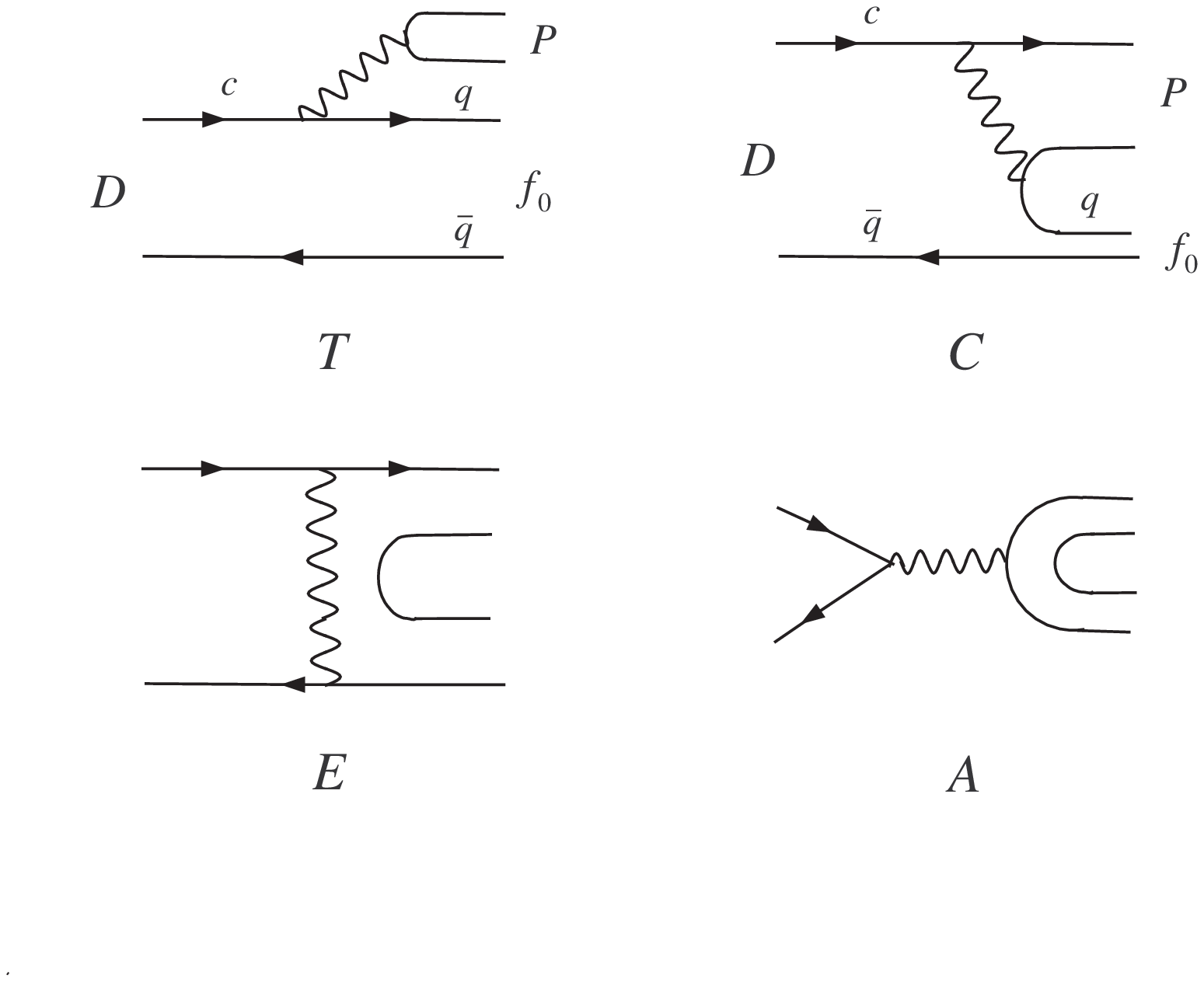,width=4in}
\vspace{-1cm}
    \caption[]{\small Topological quark diagrams for $D\to
    f_0(1370)P$ decays. The diagram $C'$ is the same as the diagram $C$ except
    for an interchange between $P$ and $f_0(1370)$.
    }
\end{figure}

Hadronic charm decays are conventionally studied within the
framework of generalized factorization in which the hadronic decay
amplitude is expressed in terms of factorizable terms multiplied
by the {\it universal} (i.e. decay process independent) effective
parameters $a_i$ that are renormalization scale and scheme
independent. In this approach, the quark-graph amplitudes read
 \be \label{qdafact}
 T_u &=& {G_F\over\sqrt{2}}a_1\,f_\pi
 F_0^{Df_0}(m_\pi^2)(m_D^2-m_{f_0}^2), \non \\
 T_s &=& {G_F\over\sqrt{2}}a_1\,f_\pi
 F_0^{D_sf_0}(m_\pi^2)(m_{D_s}^2-m_{f_0}^2), \non \\
 C_u &=& {G_F\over\sqrt{2}}a_2\,f_K
 F_0^{Df_0}(m_K^2)(m_D^2-m_{f_0}^2),  \non \\
 C'_s &=& {G_F\over\sqrt{2}}a_2\,f_{f_0}
 F_0^{D\pi}(m_{f_0}^2)(m_D^2-m_\pi^2),  \\
 E_q &=& {G_F\over\sqrt{2}}a_2\,f_DF_0^{0\to f_0^{q\bar q}\bar
 K^0}(m_D^2)(m_{f_0(1370)}^2-m_K^2), \non \\
 A_q &=& {G_F\over\sqrt{2}}a_2\,f_DF_0^{0\to f_0^{q\bar q}\pi^+}
 (m_D^2)(m_{f_0(1370)}^2-m_\pi^2), \non
 \en
where the form factor $F_0$ is defined in \cite{BSW} and the
typical values of $a_i$ in charm decays are $a_1=1.15$ and
$a_2=-0.55$\,. For $f_0(1370)$, its decay constant $f_{f_0(1370)}$
is zero owing to charge conjugation invariance or conservation of
vector current \cite{Diehl}. This means that the amplitude $C'_s$
vanishes under the factorization approximation.

In Eq. (\ref{qdafact}) the annihilation form factor $F_0^{0\to
f_0P }(m_D^2)$ is expected to be suppressed at large momentum
transfer, $q^2=m_D^2$, corresponding to the conventional helicity
suppression. Based on the helicity suppression argument, one may
therefore neglect short-distance (hard) $W$-exchange and
$W$-annihilation contributions. However, as stressed in
\cite{a1a2charm}, weak annihilation does receive long-distance
contributions from nearby resonances via inelastic final-state
interactions from the leading tree or color-suppressed amplitude.
The effects of resonance-induced FSIs can be described in a model
independent manner and are governed by the masses and decay widths
of the nearby resonances. Indeed, the weak annihilation
($W$-exchange $E$ or $W$-annihilation $A$) amplitude for $D\to PP$
decays has a sizable magnitude comparable to the color-suppressed
internal $W$-emission amplitude $C$ with a large phase relative to
the tree amplitude $T$ [see Eq. (\ref{PP})].

In the $q\bar q$ description of $f_0(1370)$, it follows from Eq.
(\ref{f0w.f.}) that
 \be
 F_0^{D^0f_0}={1\over\sqrt{2}}\cos\theta\,F_0^{D^0f_0^{u\bar u}}, \qquad
 F_0^{D^+f_0}={1\over\sqrt{2}}\cos\theta\,F_0^{D^+f_0^{d\bar d}},
 \qquad
 F_0^{D_s^+f_0}=\sin\theta\,F_0^{D_s^+f_0^{s\bar s}},
 \en
where the superscript $q\bar q$ denotes the quark content of $f_0$
involved in the transition. In the limit of SU(3) symmetry,
$F_0^{D^0f_0^{u\bar u}}=F_0^{D^+f_0^{d\bar
d}}=F_0^{D^+_sf_0^{s\bar s}}$ and hence
 \be
 F_0^{D^0f_0}=F_0^{D^+f_0}={1\over\sqrt{2}}\,F_0^{D_s^+f_0}\,\cot\theta.
 \en
Consequently, under the factorization approximation one has
$T_s=\sqrt{2}\,T_d\tan\theta$, a relation valid in the more
general diagrammatic approach.

Since
 \be
 {\Gamma(D_s^+\to f_0(1370)\pi^+)\over \Gamma(D^+\to
 f_0(1370)\pi^+)} ={\B(D_s^+\to f_0(1370)\pi^+)\over \B(D^+\to
 f_0(1370)\pi^+)}\,{\tau(D^+)\over\tau(D_s^+)},
 \en
it follows from Eqs. (\ref{BRdata}) and (\ref{qda}) that
 \be \label{Ts/Td}
 \left|{T_s+A_{u,d}\over T_d-C'_s+A_{u,d}}\right|_{D\to f_0(1370)P}=(0.76\pm
 0.24)\left({\B(f_0(1370)\to K^+K^-)\over \B(f_0(1370)\to
 \pi^+\pi^-)}\right)^{1/2},
 \en
where the charmed meson lifetimes are taken from \cite{PDG}. Let
us consider two extreme cases: (i) the $W$-annihilation term
vanishes, and (ii) $f_0(1370)$ is purely a $n\bar n$ state so that
$T_s=0$.

To proceed we will take $C'_s=0$ as suggested by the factorization
approach.  In the case of a vanishing $W$-annihilation,
$A_{u,d}=0$. Hence, the left hand side of Eq. (\ref{Ts/Td})
becomes $\sqrt{2}\,|\!\tan\theta|$. In order to estimate the
mixing angle we use the measurement of $R\equiv\Gamma(K\ov
K)/\Gamma(\pi\pi)=0.46\pm0.15\pm0.11$ \cite{WA102}.\footnote{A
reanalysis of the old data on the reactions $\pi^-p\to\pi^-\pi^+n$
and $\pi^+\pi^-\to K\ov K$ yields $R=1.33\pm0.67$ \cite{Bugg}.
This is inconsistent with naive expectation. First, the $\pi\pi$
phase space is larger than the $K\ov K$ one by a factor of 1.8\,.
Second, the $g_{f_0\pi\pi}$ coupling is larger than $g_{f_0K\ov
K}$ if $f_0(1370)$ is mostly $n\bar n$.} This leads to
 \be
 {\Gamma(f_0(1370)\to
 K^+K^-)\over\Gamma(f_0(1370)\to\pi^+\pi^-)}=0.35\pm0.14\,.
 \en
From Eq. (\ref{Ts/Td}) we obtain
 \be \label{theta}
 \theta=\pm (17.5^{+6.5}_{-5.9})^\circ.
 \en
This means that even in the absence of $W$-annihilation, a small
amount of the $s\bar s$ content in the $f_0(1370)$ wave function
will suffice to account for the observed rate of $D_s^+\to
f_0(1370)\pi^+$ relative to $D^+\to f_0(1370)\pi^+$.

In the other extreme case where $f_0(1370)$ is a pure $n\bar n$
state, $D_s^+\to f_0(1370)\pi^+$ can proceed only via
$W$-annihilation which includes both short-distance and
long-distance effects. Even the short-distance $W$-annihilation is
helicity suppressed, a long-distance contribution to the
topological $W$-annihilation in $D^+_s\to f_0(1370)\pi^+$ arises
from the color-allowed decay $D^+_s\to f_0(980)\pi^+$ followed by
a resonant-like rescattering as depicted in Fig. 2. Note that the
flavor wave function of $f_0(980)$ has the symbolic expression
$s\bar s(u\bar u+d\bar d)/\sqrt{2}$ \cite{Jaffe} as the light
scalars are favored to be 4-quark states (for a recent discussion,
see, e.g. \cite{ChengDSP}). The decay $D^+_s\to f_0(980)\pi^+$ has
a large branching ratio of $(1.8\pm0.3)\%$ \cite{ChengDSP}. As
discussed in \cite{a1a2charm}, Fig. 2 manifested at the hadron
level receives a $s$-channel resonant contribution from, for
example, the $0^-$ resonance $\pi(1800)$ and a $t$-channel
contribution with one-particle exchange. It follows from Eq.
(\ref{Ts/Td}) that
 \be \label{A/T}
 \left| {A_{u,d}\over T_d+A_{u,d}}\right|_{D\to f_0(1370)P}=0.45\pm 0.18\,.
 \en
The magnitude of $A/T$ depends on the its phase. Since
$W$-annihilation is expected to be dominated by the imaginary
part, we will have $|A_{u,d}/T_d|=0.50^{+0.36}_{-0.17}$ if the
relative phase between $A$ and $T$ is $90^\circ$, for example.
This means that if $f_0(1370)$ is composed of only $n\bar n$, then
one will need a very sizable $W$-annihilation to account for the
observed $D_s^+\to f_0(1370)\pi^+$ decay. However, recall that in
Cabibbo-allowed $D\to PP$ decays, the topological amplitudes given
in Eq. (\ref{PP}) lead to
 \be \label{E/T}
 \left.{A\over T}\right|_{D\to PP}=
 (0.39\pm0.20)\,e^{-i(65\pm30)^\circ}.
 \en
This indicates that although the $W$-annihilation term induced
from nearby resonances via FSIs is sizable, it is probably
unlikely that it can be big enough to satisfy the constraint
(\ref{A/T}). In reality, both external $W$-emission and
$W$-annihilation contribute to the decay and the $s\bar s$
component in $f_0(1370)$ is smaller than that implied by Eq.
(\ref{theta}).

\begin{figure}[t]
\hspace{1.0cm}
  \psfig{figure=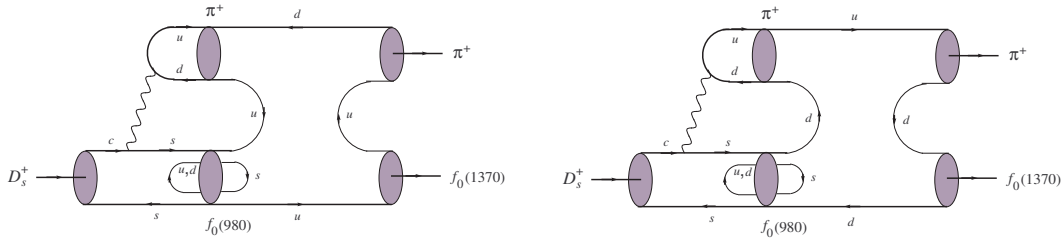,width=5.8in}
\vspace{-7.5cm}
    \caption[]{\small Contributions to $D^+_s\to f_0(1370)\pi^+$ from
    the color-allowed weak decay $D^+_s\to f_0(980)\pi^+$ followed by a
    resonant-like rescattering. This has the same topology as the
    $W$-annihilation graph. The flavor wave function of $f_0(980)$
    has the symbolic expression $s\bar s(u\bar u+d\bar d)/\sqrt{2}$.}
\end{figure}

\section{$D^0\to \lowercase{f_0(1370)}\ov K^0$ and the finite width effect}

We next turn to the decay $D^0\to f_0(1370)\ov K^0$ relative to
$D^+\to f_0(1370)\pi^+$. From Eqs. (\ref{BRdata}) and (\ref{qda})
we have
 \be \label{fK}
 \left| {C_u+E_{d,s}\over T_d-C'_s+A_{u,d}}\right|_{D\to f_0(1370)P}=(0.58\pm
 0.15){1\over\sqrt{r}},
 \en
where $r=p_c(D^0\to f_0\ov K^0)/p_c(D^+\to f_0\pi^+)$, and $p_c$
is the c.m. momentum of the final-state particles in the rest
frame of the charmed meson. However, the momentum $p_c$ in the
decay $D^0\to f_0(1370)\ov K^0$ is very sensitive to the
$f_0(1370)$ mass. For example, $p_c=0,~34,~214$ MeV and hence
$r=0,~0.083,~0.47$ for $m_{f_0}=1400,~1370,~1310$ MeV,
respectively. Therefore, when $m_{f_0}=1370$ MeV, one needs
$C/T\sim 7$ to account for the observed decay rate of $D^0\to
f_0(1370)\ov K^0$ relative to $D^+\to f_0(1370)\pi^+$, which is
certainly very unlikely. The difficulty has something to do with
the decay width of the scalar resonance which we have neglected so
far. As the decay $D^0\to f_0(1370)\ov K^0$ is marginally or even
not allowed kinematically, depending on the $f_0(1370)$ mass, it
is important to take into account the finite width effect of the
resonance. That is, one should evaluate the two-step process
$\Gamma(D^0\to f_0(1370)\ov K^0\to \pi^+\pi^-\ov K^0)$ and compare
the resonant three-body rate with experiment.

The decay rate of the resonant three-body decay is given by
 \be \label{3body}
 \Gamma(D\to SP\to P_1P_2P) &=& {1\over
 2m_D}\int^{(m_D-m_P)^2}_{(m_1+m_2)^2}{dq^2\over 2\pi}\,|\la
 SP|{\cal H}_W|D\ra|^2\,{\lambda^{1/2}(m_D^2,q^2,m_P^2)\over 8\pi m_D^2}  \non \\
 &\times& {1\over (q^2-m_S^2)^2+(\Gamma_{12}(q^2)m_S)^2}\,g^2_{SP_1P_2}
 {\lambda^{1/2}(q^2,m_1^2,m_2^2)\over 8\pi q^2},
 \en
where $\lambda$ is the usual triangluar function
$\lambda(a,b,c)=a^2+b^2+c^2-2ab-2ac-2bc$, $m_1$ ($m_2$) is the
mass of $P_1$ ($P_2$), and the ``running" or ``comoving" width
$\Gamma_{12}(q^2)$ is a function of the invariant mass
$m_{12}=\sqrt{q^2}$ of the $P_1P_2$ system and it has the
expression \cite{Pilkuhn}
 \be
 \Gamma_{12}(q^2)=\Gamma_S\,{m_S\over m_{12}}\,{p'(q^2)\over
 p'(m_S^2)},
 \en
where  $p'(q^2) = \lambda^{1/2}(q^2,m_1^2,m_2^2)/(2\sqrt{q^2})$ is
the c.m. momentum of $P_1$ or $P_2$ in the $P_1P_2$ rest frame and
$p'(m_S^2)$ is the c.m. momentum of either daughter in the
resonance rest frame. The propagator of the resonance is assumed
to be of the Breit-Wigner form.

When the resonance width $\Gamma_S$ is narrow, the expression of
the resonant decay rate can be simplified by applying the
so-called narrow width approximation
 \be
 {1\over (q^2-m_S^2)^2+m_S^2\Gamma_{12}^2(q^2)}\approx {\pi\over
 m_S\Gamma_S}\delta(q^2-m_S^2).
 \en
Noting
 \be
 \Gamma(D\to SP) = |\la SP|{\cal H}_W|D\ra|^2\, {p\over 8\pi m_D^2},
 \qquad\quad
 \Gamma(S\to P_1P_2) = g^2_{SP_1P_2}\,{p'(m_S^2)\over 8\pi m_S^2},
 \en
where $p=\lambda^{1/2}(m_D^2,m_S^2,m_P^2)/(2m_D)$ is the c.m.
three-momentum of final-state particles in the $D$ rest frame, we
are led to the ``factorization" relation
 \be \label{fact}
 \Gamma(D\to SP\to P_1P_2P)=\Gamma(D\to SP)\B(S\to P_1P_2)
 \en
for the resonant three-body decay rate.

In practice, this factorization relation works reasonably well as
long as the two-body decay $D\to SP$ is kinematically allowed and
the resonance is narrow. However, when $D\to SP$ is kinematically
barely or even not allowed, the off resonance peak effect of the
intermediate resonant state will become important. For example,
the fit fractions of $D^0\to \rho(1700)^+K^-\to \pi^+\pi^0K^-$,
$D^0\to K^*_0(1480)\ov K^0\to K^+\pi^-\ov K^0$ have been measured
by CLEO \cite{CLEO} and BaBar \cite{BaBar}, respectively. It is
clear that the on-shell decays $D^0\to \rho(1700)^+K^-$ and
$D^0\to K^*_0(1480)\ov K^0$ are kinematically not allowed and it
is necessary to take into account the finite width effect.

Since $f_0(1370)$ is broad with a width ranging from 200 to 500
MeV, {\it a priori} there is no reason to neglect its finite width
effect. For simplicity in practical calculations, we shall fix the
weak matrix element $\la SP|{\cal H}_W|D\ra$ and the strong
coupling $g_{SP_1P_2}$ at $q^2=m_S^2$ and assume that they are
insensitive to the $q^2$ dependence when the resonance is off its
mass shell. Let us define the parameter $\eta$
 \be
 \eta\equiv {\Gamma(D\to SP\to
 P_1P_2P)\over \Gamma(D\to SP)\B(S\to P_1P_2)}.
 \en
The deviation of $\eta$ from unity will give a measure of the
violation of the factorization relation (\ref{fact}). Then it has
the expression
 \be
 \eta &=& {m_S^2\over 4\pi m_D}\,{\Gamma_S\over pp'(m_S^2)}\int^{(m_D-m_P)^2}_{(m_1+m_2)^2}
 {dq^2\over q^2}\,
 \lambda^{1/2}(m_D^2,q^2,m_P^2)\lambda^{1/2}(q^2,m_1^2,m_2^2)
 \non \\ &\times& {1\over (q^2-m_S^2)^2+(\Gamma_{12}(q^2)m_S)^2}.
 \en
For $m_{f_0(1370)}=1370$ MeV and $\Gamma_{f_0(1370)}=200$ MeV (500
MeV), we find $\eta=3.8~(4.3),~0.83~(0.67),~0.89~(0.74)$ for the
decays $D^0\to f_0(1370)\ov K^0\to \pi^+\pi^-\ov K^0$, $D^+\to
f_0(1370)\pi^+\to\pi^+\pi^-\pi^+$ and $D_s^+\to
f_0(1370)\pi^+\to\pi^+\pi^-\pi^+$, respectively. It is evident
that the finite width effect of $f_0(1370)$ is very crucial for
$D^0\to f_0(1370)\ov K^0$. This also indicates that the measured
branching ratios shown in (\ref{BRdata}) are actually for resonant
three-body decays.

Let us return back to Eq. (\ref{fK}). The parameter $r$ there
should be replaced by $r=I_1/I_2$ with
 \be
 I_1 &=& \int^{(m_D-m_\pi)^2}_{4m_K^2}{dq^2\over q^2}\,\lambda^{1/2}
 (m_D^2,q^2,m_\pi^2)\,\lambda^{1/2}(q^2,m_\pi^2,m_\pi^2)\,
 {1\over (q^2-m_{f_0}^2)^2+(\Gamma_{12}(q^2)m_{f_0})^2}, \non \\
 I_2 &=& \int^{(m_D-m_K)^2}_{4m_\pi^2}{dq^2\over q^2}\,\lambda^{1/2}
 (m_D^2,q^2,m_K^2)\,\lambda^{1/2}(q^2,m_\pi^2,m_\pi^2)\,
 {1\over (q^2-m_{f_0}^2)^2+(\Gamma_{12}(q^2)m_{f_0})^2}.
 \en
Note that the lower bound of the integral $I_1$ is $4m_K^2$ rather
than $4m_\pi^2$ in order to have a real $p'(q^2)$. For the
representative values of $m_{f_0(1370)}=1370$ MeV and
$\Gamma_{f_0(1370)}=250$ MeV, we find $r=0.36$ and hence
 \be
 \left|{C_u+E_{d,s}\over T_d-C'_s+A_{u,d}}\right|_{D\to f_0(1370)P}=0.97\pm0.25\,,
 \en
which is to be compared with
 \be
 \left| {C+E\over T+A}\right|_{D\to PP}\sim 0.78\
 \en
in $D\to PP$ decays [see Eq. (\ref{PP})]. Therefore, the decay
$D^+\to f_0(1370)\pi^+\to\pi^+\pi^-\pi^+$ can be explained once
the finite width effect of $f_0(1370)$ is taken into account.

The comparison of $D^0\to f_0(1370)\ov K^0$ with $D_s^+\to
f_0(1370)\pi^+$ in principle allows one to obtain some information
on the mixing angle. However, since the relation between the
amplitudes $C_u$ and $T_s$ is unknown, it does not allow a
model-independent extraction. Finally, it should be remarked that
owing to the finite width effect, Eqs. (\ref{theta}) and
(\ref{A/T}) are slightly modified to
 \be
 \theta=\pm(18.8^{+6.8}_{-7.4})^\circ, \qquad \qquad \left| {A_{u,d}\over
 T_d+A_{u,d}}\right|_{D\to PP}=0.48\pm0.20\,.
 \en

\section{Discussion and Conclusion}
The decay $D^+\to f_0(1370)\pi^+$ receives the main contribution
from the external $W$-emission diagram via the $n\bar n$ component
of $f_0(1370)$, while $D_s^+\to f_0(1370)\pi^+$ proceeds via the
external $W$-emission through the $s\bar s$ content; both channels
receive $W$-annihilation. Assuming the absence of
$W$-annihilation, we showed in a model independent way that both
modes can be accommodated provided that
$\theta=\pm(17.5^{+6.5}_{-5.9})^\circ$. That is, even a small
$s\bar s$ component in $f_0(1370)$ can induce adequate $D_s^+\to
f_0(1370)\pi^+$ via the external $W$-emission. In the other
extreme case where $f_0(1370)$ is a pure $n\bar n$ state, it is
found that one needs a very large $W$-annihilation to explain the
decay $D_s^+\to f_0(1370)\pi^+$. Therefore, we conclude that
$f_0(1370)$ is unlikely a pure $n\bar n$ state. In reality, both
external $W$-emission and $W$-annihilation contribute to the decay
and the mixing angle is smaller than the above-mentioned value.

To extract the upper limit on the mixing angle we have employed
the experimental value of $\Gamma(K\ov K)/\Gamma(\pi\pi)$. The
uncertainty with the branching fractions of $f_0(1370)$ can be
circumvented if $D_s^+\to f_0(1370)\pi^+\to K^+K^-\pi^+$ is
measured and compared with $D^+\to f_0(1370)\pi^+\to
\pi^+\pi^-\pi^+$.

For the decay $D^0\to f_0(1370)\ov K^0$ which is barely or even
not allowed kinematically, depending on the mass of $f_0(1370)$,
it is important to take into account the finite width effect of
$f_0(1370)$. We find that it plays a crucial role on the resonant
three-body decay $D^0\to f_0(1370)\ov K^0\to\pi^+\pi^-\ov K^0$.

\vskip 2.0cm \acknowledgments This work was supported in part by
the National Science Council of R.O.C. under Grant No.
NSC91-2112-M-001-038.

%%%%%%%%%%%%%%%%%%%%%%%%%%%%%%%%%%%%%%%%%%%%%%%%%%%%%%%%

\end{document}